# Bridging Scales in Black Hole Accretion and Feedback: Magnetized Bondi Accretion in 3D GRMHD

Hyerin Cho (조혜린)[1,2], Ben S. Prather[3], Ramesh Narayan[1,2], Priyamvada Natarajan[2,4,5], Kung-Yi Su[1,2], Angelo Ricarte[1,2], and Koushik Chatterjee[1,2]
[1] Center for Astrophysics | Harvard & Smithsonian, 60 Garden Street, Cambridge, MA 02138, USA
[2] Black Hole Initiative at Harvard University, 20 Garden Street, Cambridge, MA 02138, USA
[3] CCS-2, Los Alamos National Laboratory, P.O. Box 1663, Los Alamos, NM 87545, USA
[4] Department of Astronomy, Yale University, Kline Tower, 266 Whitney Avenue, New Haven, CT 06511, USA
[5] Department of Physics, Yale University, P.O. Box 208121, New Haven, CT 06520, USA

## Abstract

Fueling and feedback couple supermassive black holes (SMBHs) to their host galaxies across many orders of magnitude in spatial and temporal scales, making this problem notoriously challenging to simulate. We use a multi-zone computational method based on the general relativistic magnetohydrodynamic (GRMHD) code KHARMA that allows us to span 7 orders of magnitude in spatial scale, to simulate accretion onto a non-spinning SMBH from an external medium with a Bondi radius of $R_B \approx 2 \times 10^5\, GM_\bullet/c^2$, where $M_\bullet$ is the SMBH mass. For the classic idealized Bondi problem, spherical gas accretion without magnetic fields, our simulation results agree very well with the general relativistic analytic solution. Meanwhile, when the accreting gas is magnetized, the SMBH magnetosphere becomes saturated with a strong magnetic field. The density profile varies as $\sim r^{-1}$ rather than $r^{-3/2}$ and the accretion rate $\dot{M}$ is consequently suppressed by over 2 orders of magnitude below the Bondi rate $\dot{M}_B$. We find continuous energy feedback from the accretion flow to the external medium at a level of $\sim 10^{-2} \dot{M}c^2 \sim 5 \times 10^{-5} \dot{M}_B c^2$. Energy transport across these widely disparate scales occurs via turbulent convection triggered by magnetic field reconnection near the SMBH. Thus, strong magnetic fields that accumulate on horizon scales transform the flow dynamics far from the SMBH and naturally explain observed extremely low accretion rates compared to the Bondi rate, as well as at least part of the energy feedback.

*Unified Astronomy Thesaurus concepts:* Accretion (14); Active galactic nuclei (16); Bondi accretion (174); Schwarzschild black holes (1433); Supermassive black holes (1663); Hydrodynamical simulations (767)

## 1. Introduction

Most nearby galaxies are found to harbor central supermassive black holes (SMBHs), whose masses are correlated with the properties of the stellar components of their hosts (e.g., Magorrian et al. 1998; Ferrarese & Merritt 2000; Gebhardt et al. 2000; Kormendy & Ho 2013). However, the details of how gas flows into the galactic nucleus from large cosmic distances and how the SMBH in turn imparts feedback into the galaxy remain coupled unresolved problems.

Due to limitations in computational power and resolution, typically there has been a partitioning of scales in tackling the SMBH feeding and the feedback problem. In galaxy-scale simulations, the gas flows from kiloparsec scales and the coevolution of an SMBH and a galaxy are explicitly modeled (e.g., Sijacki et al. 2015; Rosas-Guevara et al. 2016; Weinberger et al. 2018; Ricarte et al. 2019; Ni et al. 2022; Wellons et al. 2023). However, even in idealized simulations of an isolated galactic nucleus, prescriptions need to be adopted as sub-grid models to include accretion and feedback (e.g., Li & Bryan 2014; Fiacconi et al. 2018; Anglés-Alcázar et al. 2021; Talbot et al. 2021; Weinberger et al. 2023) usually at a scale $\lesssim 10$ pc, which translates to $\sim 3 \times 10^4 r_g$ for the $\sim 6.5 \times 10^9\, M_\odot$ SMBH in M87; here, $r_g = GM_\bullet/c^2$ is the gravitational radius, where $M_\bullet$ is the SMBH mass. Meanwhile, probing smaller scales, general relativistic magnetohydrodynamic (GRMHD) simulations focus on the inner few tens or hundreds of $r_g$, self-consistently tracing gas flows and feedback on these limited spatial scales (e.g., Gammie et al. 2003; Tchekhovskoy et al. 2011; Porth et al. 2019; Chatterjee et al. 2023). Typically, GRMHD simulations use idealized settings and disregard the larger-scale cosmological environment. Bridging these vastly different scales while self-consistently following gas flows and incorporating both the GRMHD effects and cosmological evolution remains challenging.

Recently, some attempts to connect these disparate scales have been made using nested simulations where each smaller-scale simulation is initialized from the next larger-scale simulation (e.g., Hopkins & Quataert 2010; Ressler et al. 2020; Guo et al. 2023), or by adopting Lagrangian hyper-refinement methods (e.g., Anglés-Alcázar et al. 2021; Hopkins et al. 2023), or by pushing out the simulation regime to larger scales (e.g., Lalakos et al. 2022; Kaaz et al. 2023). While successful in studying the process of accretion, feedback from the SMBH has not been followed out to galaxy scales in prior work because either the communication between scales is only directed inward or because of the inability to include the entire galaxy and the black hole.

We employ here a multi-zone computational method that represents a first attempt to not only follow accretion flows but also to trace feedback across 7 orders of magnitude from the event horizon to galactic scales. With this method, we study purely hydrodynamic (HD) Bondi accretion as well as the magnetized Bondi problem.

This paper is outlined as follows: in Section 2, we give an overview of the numerical scheme that captures a wide dynamic range simultaneously. In Section 3, we present a purely HD simulation and test our numerical scheme by comparing it with the general relativistic (GR) analytical solution. In Section 4, we include magnetic fields for a more realistic representation of the environment of SMBHs and study how Bondi accretion is modified. We discuss our findings of feedback via convection in Section 5 and summarize the conclusions of our study in Section 6. Additional details are presented in a set of short appendices: various GRMHD quantities are defined in Appendix A, the details of the simulation setup are outlined in Appendix B, and a study of resolutions and initial conditions is presented in Appendix C.

Throughout, we adopt units where $GM_\bullet = c = 1$, so our unit of length is $r_g = 1$ and time is $t_g \equiv r_g/c = 1$. Gas temperature $T$ is a dimensionless quantity that is normalized by $\mu c^2/k_B$, where $\mu$ is the mean molecular weight and $k_B$ is the Boltzmann constant; thus, the gas pressure is $p_g = \rho T$ with $\rho$ the density. We denote the time average of a quantity $X$ as $\overline{X}$ and the density-weighted, time-, and shell-averaged value of $X$ as $\langle X \rangle$ (Appendix A). Finally, the Bondi radius is $R_B \equiv GM_\bullet/c_{s,\infty}^2$, where $c_{s,\infty} = \sqrt{\gamma_{\rm ad} T_\infty}$ is the sound speed in the external medium far outside $R_B$, $T_\infty$ is the temperature there, and $\gamma_{\rm ad}$ is the gas adiabatic index. The freefall time at the Bondi radius is $t_B \equiv (R_B/r_g)^{3/2} t_g$. For the models presented here, $\gamma_{\rm ad} = 5/3$, $T_\infty = 3.4 \times 10^{-6}$ ($\approx 2 \times 10^7$ K), $R_B = 1.8 \times 10^5 r_g$, and we consider a non-spinning SMBH (Schwarzschild metric).

## 2. Numerical Methods

Our numerical scheme utilizes the GRMHD code KHARMA,[6] a performance-portable C++ implementation based on iharm3D (Prather et al. 2021); iharm3D is itself an extension of HARM, an efficient second-order conservative finite-volume scheme for solving magnetohydrodynamic (MHD) equations on Eulerian meshes in stationary curved spacetimes (Gammie et al. 2003). KHARMA offers a more flexible, portable, and scalable implementation suitable for multiple uses, by leveraging the Parthenon Adaptive Mesh Refinement Framework and the Kokkos programming model (Trott et al. 2022; Grete et al. 2023).

The approach adopted thus far to handle a wide range of spatial and temporal scales has been to run a series of nested simulations sequentially, starting from the outside and proceeding down to the BH (e.g., Ressler et al. 2020; Guo et al. 2023). This setup is appropriate for studying the feeding of the BH from the external galactic medium (or stellar winds), but it cannot tackle feedback from the BH to the galaxy. To handle both processes, we run nested simulations from radii of $r \gg R_B$ down to $r_g$ as well as from $r_g$ out to $r \gg R_B$. In addition, we cycle in and out hundreds of times to ensure self-consistent, two-way communication between the BH and the external medium.

We use a grid based on spherical coordinates and split the simulation volume between $r = r_g$ and $r \gg R_B$ into a set of $n$ overlapping annuli evenly spaced in $\log(r)$, each of which we label as zone-$i$. For $i \in \{0, 1,...,(n-1)\}$, zone-$i$ extends from an inner radius $r_{i,\rm in} = 8^i r_g$ to an outer radius $r_{i,\rm out} = 8^{i+2} r_g$. We start by initializing and running zone-$(n-1)$, the largest annulus, for some fraction of its freefall time. We then simulate zone-$(n-2)$, inheriting the final state of zone-$(n-1)$ as the initial condition over half the new domain. This continues inward to zone-0 and back out in a sort of V cycle. Our method can be thought of as defining a global domain, but *pausing* everything outside the particular active region, shifting the active region up and down the range of scales a large number of times (in our fiducial MHD run, the V cycle was traversed 254 times) until the full system converges to a steady state.[7]

## 3. HD Bondi Accretion

We first test the accuracy of our method by numerically simulating HD spherical Bondi (1952) accretion on a Schwarzschild BH; a relativistic analytical solution is available for this problem (Michel 1972; Shapiro & Teukolsky 1983). We use a total of $n = 7$ annuli with an outer radius of $8^8 \sim 2 \times 10^7 r_g$ and a total runtime of $t_{\rm run}(R_B) \sim 9 t_B$ ($\sim$0.7 Myr for M87) at the Bondi radius. We confirmed convergence by checking that the results at $t_{\rm run} = 9 t_B$ remain unchanged even when we run up to $t_{\rm run} > 56 t_B$. The resolution is $128^3$ per annulus giving a total simulation resolution of $512 \times 128 \times 128$. Details of the setup are presented in Appendix B.1.

Figures 1(a) and (b) show the radial profiles of $t$, $\theta$, $\varphi$-averaged gas density $\rho$, and temperature $T$ (red lines). The simulation results show excellent agreement with the GR Bondi analytic solution (thick gray line) over 7 orders of magnitude in radius. Further, the time-averaged accretion rate $\overline{\dot{M}}(r)$ shown in Figure 1(c) matches the analytical Bondi accretion rate $\dot{M}_B$ very well, $\overline{\dot{M}}/\dot{M}_B = 1$, across all radii within $R_B$, confirming a steady state. We note that Guo et al. (2023) have previously simulated Bondi accretion under Newtonian gravity spanning 5 orders of magnitude. Here we have successfully reproduced the GR version over a larger range of scales.

Analogous to $\overline{\dot{M}}(r)$ we can define a time-averaged energy inflow rate toward the BH, $\overline{\dot{E}}(r)$ (Equation (A5)), but this is not useful for our purposes since it includes the rest mass energy of the accreted gas. Instead, we consider $\dot{E}_{\rm net}(r) \equiv \overline{(\dot{M} - \dot{E})}$ (Equation (A6)), which removes the rest mass contribution and is flipped in sign such that it is the net rate of outflow of energy (i.e., feedback) from the BH to large radii. Normalizing by $\overline{\dot{M}}_{10} \equiv \overline{\dot{M}}(r = 10 r_g)$, we then obtain a feedback efficiency $\eta^{\rm tot}$ (Equation (A7)),

$$\eta^{\rm tot} \equiv \overline{(\dot{M} - \dot{E})}/\overline{\dot{M}}_{10}. \quad (1)$$

Following Appendix A, and noting that there is no magnetic field and the system is spherically symmetric, $\eta^{\rm tot} \approx 1 + (1 + \gamma_{\rm ad} u/\rho) u_t$. At large radii, $u \to \rho T_\infty/(\gamma_{\rm ad} - 1)$ and $u_t \to -1$; hence,

$$\eta^{\rm tot} \to -\gamma_{\rm ad} T_\infty/(\gamma_{\rm ad} - 1) = -1.5/R_B \approx -8.4 \times 10^{-6}. \quad (2)$$

The agreement of the simulation results in Figure 1(d) with this theoretical expectation is excellent.

## 4. Magnetized Bondi Accretion

Accretion flows in galactic nuclei involve magnetized plasma, and it has been long known that magnetic fields strongly perturb the energetics (Shvartsman 1971; Meszaros 1975) and dynamics (Igumenshchev & Narayan 2002; Pen et al. 2003; Lalakos et al. 2022) of Bondi accretion. Apart from a toy model with a purely radial magnetic field and a non-spinning BH, where no dynamical interaction occurs between the fluid and the magnetic field (Gammie et al. 2003), there is no known analytical solution to the magnetized accretion problem. It can be studied only via numerical simulations. Our computational technique is well suited for this.

---

[6] https://github.com/AFD-Illinois/kharma

[7] A few details of the numerical method are given in Appendix B. A complete description is deferred to a future paper.

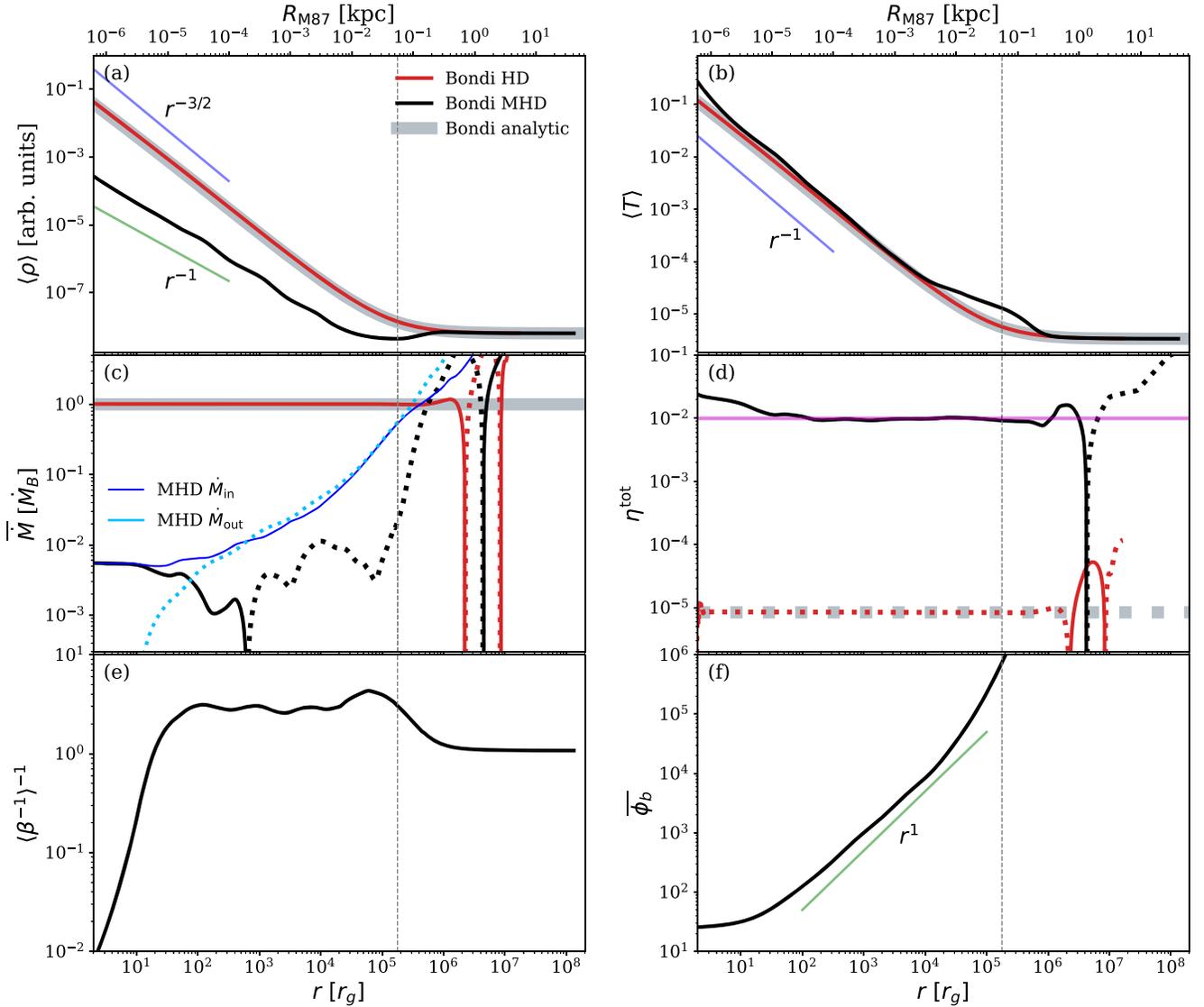

**Figure 1.** Time-averaged radial profiles of selected quantities for our HD (red) and MHD (black) Bondi accretion simulations with $R_B = 1.8 \times 10^5 \, r_g$ (gray vertical dashed line). In this and forthcoming plots, negative values are plotted with dotted rather than solid lines. While most naturally expressed in gravitational units (lower x-axis), we also present distances scaled to M87 (upper x-axis). Where available, analytic solutions of relativistic Bondi accretion are plotted with thick gray lines. Magnetic fields lead to a shallower density slope, a dramatically suppressed accretion rate, and positive energy feedback.

To simulate the MHD problem, we use the same external density and temperature as adopted for the HD case (Section 3), but now we include an initial magnetic field in the z-direction with plasma-$\beta \equiv p_g/p_b \approx 1$, where $p_b$ is the magnetic pressure. Our choice of $\beta$ is motivated by the Milky Way interstellar medium (ISM; e.g., Ferrière 2020; Guerra et al. 2023), but we have verified that the results remain unchanged if we set the initial $\beta \approx 10$ (Appendix C). We use $n = 8$ zones, which allows us to probe out to $r \sim 10^8 r_g$ (and reach a steady state out to nearly $10^7 r_g$). Our effective resolution is $576 \times 128 \times 128$ over the whole domain. Additional details of the setup are given in Appendix B.2.

Unlike the laminar flow that we find in the HD simulation, the MHD simulation gives a dynamic and turbulent flow that is driven by magnetic stresses and reconnection. We need to time average the results to obtain meaningful steady-state profiles. The black lines in Figure 1 show time-averaged radial profiles of a number of quantities after the simulation was run for a physical duration of $t_{\rm run}(R_B) \sim 9 t_B$ at the Bondi radius. The temperature profile is fairly similar to the HD Bondi case (compare the black and red lines), but the density profile is significantly modified due to the presence of the magnetic field. The radial scaling is now $\rho \propto r^{-1}$, which is similar to the slope reported in several previous GRMHD simulations of rotating accretion flows (e.g., Ressler et al. 2020; Chatterjee & Narayan 2022) and even some hydrodynamical simulations (Guo et al. 2023). It is also consistent with observational constraints in M87* and Sgr A* from X-ray and Event Horizon Telescope observations (e.g., Chatterjee & Narayan 2022).

There is a curious dip in the density and a corresponding bump in the temperature near the Bondi radius, which we explain in Section 5.

Despite running the simulation for an effective time of $\sim 9 t_B$, we were unable to obtain a reliable measurement of $\overline{\dot{M}}$ at $r \gtrsim 10^2 \, r_g$ as can be seen in Figure 1(c), unlike the HD case. This is because the medium exhibits violent convective turbulence (Section 5), which produces large random velocities that overwhelm the mean accretion velocity. To illustrate the difficulty, we show the average mass inflow rate $\dot{M}_{\rm in}$ (solid blue) and mass outflow rate $\dot{M}_{\rm out}$ (dotted cyan), defined in Appendix A, as a function of radius. With increasing distance from the BH, both rates become much larger than the net $\overline{\dot{M}}$, which is the difference between the two very similar numbers. It is therefore challenging to calculate $\overline{\dot{M}}$. Here, we restrict ourselves to smaller radii, where we believe our $\overline{\dot{M}}$ estimates are reliable. At $r = 10 \, r_g$,[8] we find $\overline{\dot{M}}_{10} \approx 0.005 \dot{M}_{\rm B}$, i.e., the accretion rate is suppressed by more than a factor of 100 relative to the HD Bondi case. This estimate is consistent with the observed accretion rate of $\sim 10^{-3}$–$10^{-2} \dot{M}_B$ in M87* (when comparing the BH accretion rate $\dot{M} \sim 10^{-3} \, M_\odot \, {\rm yr}^{-1}$ in the Event Horizon Telescope Collaboration et al. (2021) with the Bondi accretion rate $\dot{M}_B \sim 0.2 \, M_\odot \, {\rm yr}^{-1}$ in Russell et al. (2015)). It is also consistent with the observed accretion rate for Sgr A* $10^{-4} - 10^{-2} \dot{M}_B$ (Wang et al. 2013).

We plot the time-averaged energy efficiency $\eta^{\rm tot}(r)$ (Equation (1)) from the simulation shown in Figure 1(d). There are two striking features. First, the radial profile of $\eta^{\rm tot}$ is nearly constant (ideally it should be perfectly constant), which indicates that the simulation energetics have converged well.[9] Second, $\eta^{\rm tot}$ is positive, i.e., energy flows out from the accretion flow to the external medium. In other words, the simulation exhibits bona fide feedback. It is worth emphasizing that we are able to study feedback only because our numerical technique is specifically designed to allow two-way communication between the BH and the external medium (Section 2). The amount of feedback energy is $\sim 10^{-2} \overline{\dot{M}}_{10} c^2 \sim 5 \times 10^{-5} \dot{M}_B c^2$, which is lower than the values adopted in cosmological simulations $\gtrsim 10^{-3} \dot{M}_B c^2$ (e.g., Booth & Schaye 2009; Weinberger et al. 2018; Davé et al. 2019; Ni et al. 2022). However, the feedback in our runs operates without BH spin and without net rotation in the accreting gas.[10] We therefore speculate that it represents the minimum feedback that any hot quasi-spherical magnetized accretion flow will produce. The actual feedback will likely be larger when more favorable conditions are present (e.g., a spinning BH).

The radial profiles of two magnetic field related quantities are shown in the bottom panels of Figure 1: the plasma $\beta$, and the normalized enclosed magnetic flux $\phi_b$ (Tchekhovskoy et al. 2011),

$$\phi_b(r) \equiv \sqrt{\pi / \overline{\dot{M}}_{10}} \iint |B^r| \sqrt{-g} \, d\theta \, d\varphi. \quad (3)$$

We average the inverse of $\beta$ for stability, which is a commonly used technique. The profile we obtain has a flat plateau with a nearly constant value of $\beta \sim 3$ between $r = 10^2$–$10^5$. Over this extended volume, the field is in equipartition with the gas, and the magnetic flux $\phi_b$ varies $\propto r$, indicating that the organized poloidal field varies as $B_{\rm pol} \propto r^{-1}$. The pressure in the organized field is also in equipartition: $B_{\rm pol}^2 \propto \rho T \propto r^{-2}$. We find that for radii $r < 10^2$, $\beta$ declines rapidly to well below unity, and $\phi_b$ saturates at $\sim 30$. Our results are very reminiscent of the magnetically arrested disk (MAD) model of MHD accretion (Igumenshchev et al. 2003; Narayan et al. 2003; Tchekhovskoy et al. 2011, see also Bisnovatyi-Kogan & Ruzmaikin 1974). In previous work, Ressler et al. (2020) found that the entire stellar wind-fed accretion flow in their model of Sgr A* approached the MAD state. In our case, MAD accretion extends from the Bondi radius $R_B \approx 1.8 \times 10^5 \, r_g$ down to the BH horizon at $2 \, r_g$. One minor difference is that $\phi_b \sim 30$ at the BH horizon in our simulation, whereas typical GRMHD simulations that consider orbiting gas (instead of spherical infall) find $\phi_b \sim 50$ for a non-spinning BH (Tchekhovskoy et al. 2012; Narayan et al. 2022).

### 5. Feedback via Reconnection-driven Convection

As noted, our simulation permits us for the first time to study the coupled BH feeding and feedback problem. Following the time evolution of the BH mass accretion rate $\dot{M}$ (in units of the Bondi rate $\dot{M}_B$) and the magnetic flux parameter $\phi_b$ at the horizon (top panel of Figure 2) reveals large amplitude fluctuations in both quantities around a fairly stable mean value, $\overline{\dot{M}}_{10} \sim 0.005 \dot{M}_B$ and $\overline{\phi}_b \sim 30$. The timescale of the largest fluctuations is a few Bondi times, as expected for feeding from the Bondi radius. The cause of the fluctuations is violent turbulence, evidence for which can be seen in the multi-scale snapshots in the middle and lower rows of Figure 2.

To investigate the mechanism driving the turbulence, we explore the physics of the outward flow of energy. The black curves in the upper panels of Figure 3 show the radial profile of the efficiency parameter $\eta^{\rm tot}$. The other curves show individual physical components: $\eta^{\rm fl}$ (blue, left top panel), the energy carried in the fluid (as opposed to the magnetic field), $\eta^{\rm adv}$ (green, middle), the advected part of the fluid energy, and $\eta^{\rm conv}$ (red, right), the part of the fluid energy transport that can be attributed to convection. The definitions of these components are given in Appendix A and are identical to those used by Ressler et al. (2021).

At all radii $r \lesssim R_B$, we find that $\eta^{\rm fl}$ very closely matches $\eta^{\rm tot}$, i.e., essentially all the energy flux is carried by the fluid, reminiscent of the convection-dominated accretion model of Narayan et al. (2000) and Quataert & Gruzinov (2000), while the electromagnetic contribution is negligible. The electromagnetic energy transport would likely become important if the BH had spin, since field lines near the horizon would be frame dragged and produce a relativistic jet. In a smaller-scale simulation with $R_B = GM_\bullet / c_{s,\infty}^2 = 100$ and a spinning BH ($a_* = 0.9375$), Ressler et al. (2021) found that the electromagnetic $\eta$ exceeded $\eta^{fl}$. For the non-spinning case here, we find the opposite.

---

[8] It is customary in GRMHD simulations not to measure $\dot{M}$ at the horizon, where numerical floors might affect the fidelity of the computed density, but to measure it at a larger radius. We determined that $r = 10 r_g$ is a safe choice for our work.
[9] The simulations indicate a factor of $\sim 2$ increase in $\eta^{\rm tot}$ close to the BH. We believe this is an artifact, caused by the need to apply numerical floors near the BH horizon.
[10] Because of turbulent fluctuations, the gas at each radius tends to have a random amount of rotation, at the level of a few percent of Kepler. But the rotation is too weak to have any dynamical effect

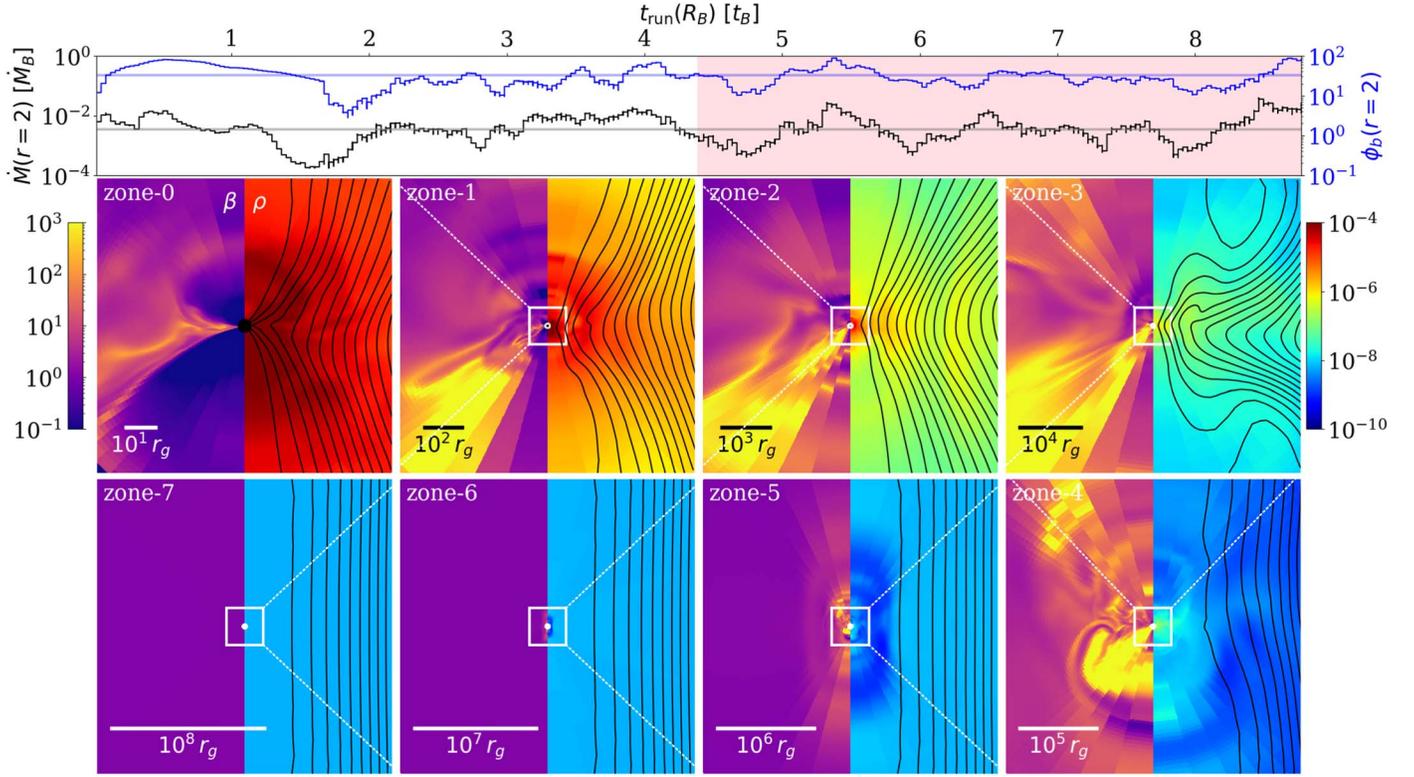

**Figure 2.** Top: accretion rate (black) and magnetic flux parameter (blue) as a function of time for our MHD simulation. The mean accretion rate of $\overline{\dot{M}}_{10} \sim 5 \times 10^{-3} \dot{M}_B$ and the mean magnetic flux at the horizon $\overline{\phi}_b \sim 30$ are plotted in gray and blue horizontal lines, respectively. We average values over the second half of the simulations (pink background), when the simulation has reached a steady state. Bottom: slices of $\beta$ (left) and $\rho$ (right) in our simulation, spanning 8 orders of magnitude in spatial scale. Lines of constant magnetic flux representing $\varphi$-averaged magnetic fields are overlaid in black.

When we divide $\eta^{fl}$ into an advective part $\eta^{adv}$ and a convective part $\eta^{conv}$, we find that convection dominates by a substantial margin between $r = r_g$ and $\sim 0.1 R_B$; energy transport is principally via convection. From the work of Igumenshchev & Narayan (2002), the convection is initiated by magnetic reconnection near the BH, which transforms the dissipated magnetic energy into thermal energy (entropy) and kinetic energy (Ripperda et al. 2022). Convective turbulence develops and extends all the way to $r \sim 0.1 R_B$, carrying the dissipated energy outward. Note that our definition of $\eta^{conv}$ closely follows that given by Ressler et al. (2021). However, it is possible that some fraction of what we call convection corresponds to a violently varying wind (e.g., Yuan et al. 2015). Interestingly, although magnetic reconnection drives the convection, the field does not transport the energy because field lines remain largely fixed in radius in the MAD state.

The bottom three panels in Figure 3 present detailed information on the fluid, advective, and convective energy fluxes, $F_E^{fl}$, $F_E^{adv}$, and $F_E^{conv}$ (defined in Appendix A), in the logarithmic radius—polar angle ($\log_{10} r$ - $\theta$) plane. The advective flux carries energy toward the BH near the midplane but away from the BH near the poles. When integrated over $\theta$, the two zones nearly cancel. On the other hand, convection transfers energy outward at all $\theta$ and overwhelms the advective energy at all $r \lesssim 0.1 R_B$ so that the net fluid flux $F_E^{fl}$ is outward everywhere. While Ressler et al. (2021) concluded that convection was inefficient in their spinning BH setup, we find exactly the opposite for our non-spinning BH. Pen et al. (2003) found energy inflow rather than outflow in their nonrelativistic simulation, which they termed *frustrated convection*. The difference from our result might be because their magnetic field did not reach the saturated MAD limit and/or their spatial dynamic range was only a few tens.

Finally, we see in Figure 3 that the convective energy transport ceases at $R_B$, as expected since convection requires gravity and buoyancy, while gravity becomes subdominant for $r > R_B$. Thus, in our model, the feedback energy carried outward by convection is dumped near the Bondi radius and has nowhere to go. This explains the bump in the temperature profile of the MHD run near the Bondi radius shown in Figure 1, and the corresponding dip in density (required by pressure balance).

## 6. Summary and Conclusions

We have simulated general relativistic accretion on a non-spinning SMBH using a numerical technique that allows us to maintain two-way communication across the entire range of spatial scales from the BH horizon to beyond the Bondi radius, $R_B \approx 1.8 \times 10^5 r_g$. For spherical accretion without magnetic fields, we find perfect agreement with the GR analytical Bondi solution, i.e., for $r < R_B$, the density and temperature of the simulated gas scale as $\rho \propto r^{-3/2}$ and $T \propto r^{-1}$, respectively, as expected, and the measured accretion rate also matches the analytical Bondi accretion rate $\dot{M}_B$. The thermal energy of the external gas, which is $1.5/R_B \approx 10^{-5}$ (Equation (2)) of the rest mass energy, is advected into the BH and there is no feedback in the opposite direction.

Adding a magnetic field of realistic strength to the external gas changes the results drastically. The density slope inside $R_B$

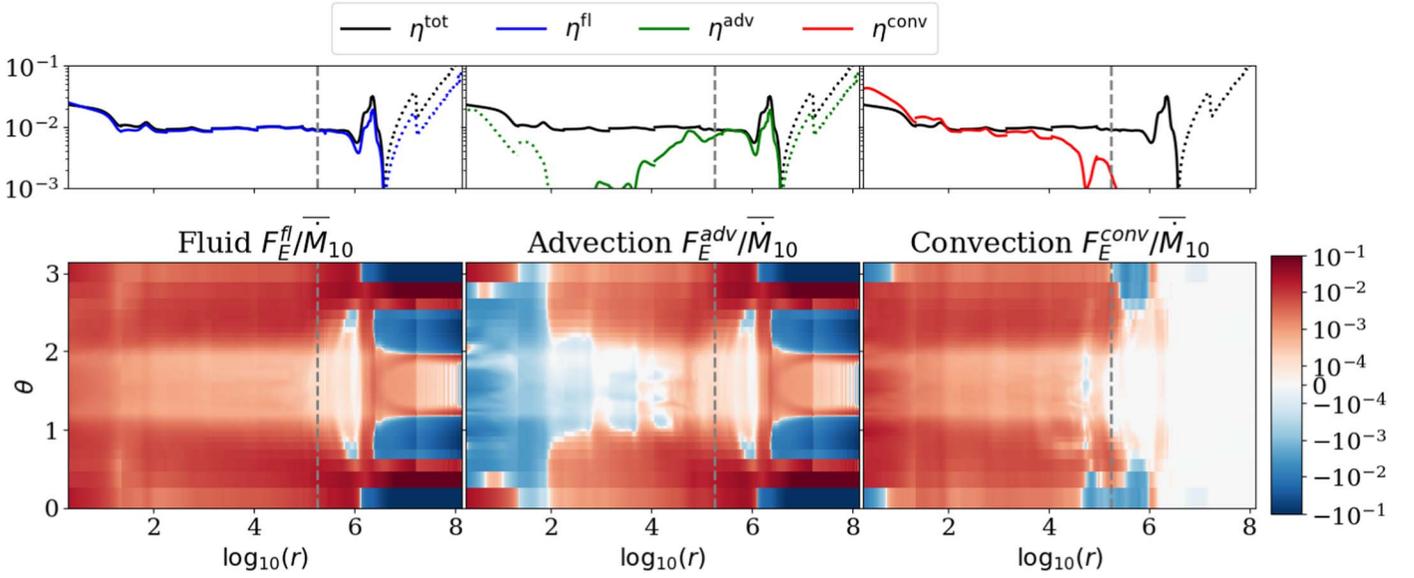

**Figure 3.** Top: dissecting contributions to the total energy outflow ($\eta^{\rm tot}$), including the fluid contribution ($\eta^{\rm fl}$), the advective outflow efficiency ($\eta^{\rm adv}$), and the convective outflow efficiency ($\eta^{\rm conv}$). Since $\eta^{\rm fl} \approx \eta^{\rm tot}$, the electromagnetic field does not contribute much directly to the energy outflow. Instead, the energy is transported primarily via convection until near the Bondi radius (gray vertical dashed line). Positive (negative) values are shown in solid (dotted) lines. Bottom: azimuthally averaged energy fluxes in the $r$, $\theta$ plane, normalized by the accretion rate at 10 $r_g$. Outflowing energy is shown in red and inflowing energy in blue.

changes to $\rho \propto r^{-1}$, and the mass accretion rate $\dot{M}$ into the BH drops dramatically to $\approx 5 \times 10^{-3} \dot{M}_{\rm B}$, consistent with observations of M87* and Sgr A*. Equipartition of magnetic and gas pressure is achieved ($\beta \sim 3$) over the radius range of $10^2 r_g \lesssim r \lesssim R_{\rm B}$, and the magnetic pressure becomes much more dominant close to the BH. Except for the lack of net rotation, the solution closely resembles the MAD model.

Significantly, the magnetized simulation shows net energy feedback from the accretion flow to the external medium at a rate of $\approx 10^{-2} \dot{M} c^2$, i.e., with an efficiency of $\eta^{\rm tot} \sim 0.01$. Expressed in terms of the Bondi accretion rate, the energy outflow rate is $\sim 5 \times 10^{-5} \dot{M}_{\rm B} c^2$. This positive feedback occurs even in the absence of BH spin or net gas rotation. The energy outflow is not strongly collimated; it is quasi-isotropic. Advection carries energy inward in the midplane and transports energy outward near the poles, while convection transfers energy outward at all $\theta$. We find that the level of energy feedback is significantly lower than that typically invoked in cosmological simulations, so we view the feedback estimate in the present work as representing a lower limit to energy outflow. Additional ingredients, especially BH spin and gas angular momentum, are needed to study the problem more fully.

The dominant outward energy transport mechanism is convection driven by magnetic reconnection in the magnetosphere near the BH. Even though a strong magnetic field is essential for this mechanism to work, the energy is carried almost entirely in the form of thermal and kinetic energy of the gas, while the electromagnetic energy flux is negligible. Thus, the feedback mechanism seen here is different from the electromagnetically dominated outflows found in relativistic jets and rotation-driven disk winds.

These new insights are possible because our numerical technique permits two-way communication between the BH and the external medium over many decades of spatial and temporal scales. In future work, we plan to extend our investigation to the case of a spinning BH that also includes the angular momentum of the gas.

**Acknowledgments**

The authors H.C., R.N., P.N., K.S., A.R., and K.C. acknowledge support from the Gordon and Betty Moore Foundation and the John Templeton Foundation via grants to the Black Hole Initiative at Harvard University. This work used Delta at the University of Illinois at Urbana-Champaign through allocation PHY230079 from the Advanced Cyberinfrastructure Coordination Ecosystem: Services & Support (ACCESS) program, which is supported by National Science Foundation grants Nos. 2138259, 2138286, 2138307, 2137603, and 2138296. This research used resources provided to B.P. by the Los Alamos National Laboratory Institutional Computing Program, which is supported by the U.S. Department of Energy National Nuclear Security Administration under Contract No. 89233218CNA000001. We thank the referee for a thorough reading of the manuscript and constructive comments that have helped improve it.

## Appendix A
## GRMHD Primer and Definitions

Grid-based GRMHD codes such as KHARMA used in this work evolve a magnetized fluid in a stationary spacetime (often a BH) described by a spacetime metric $g_{\mu\nu}$. We focus on a non-spinning BH, and hence use the Schwarzschild metric. The magnetized fluid is treated within the ideal MHD approximation with the stress-energy tensor $T^\mu_\nu$ written as (see Anile 1989; Komissarov 1999; Gammie et al. 2003 for details)

$$T^\mu_\nu = (\rho + \gamma_{\rm ad} u + b^2) u^\mu u_\nu + [p_g + (b^2/2)]\delta^\mu_\nu - b^\mu b_\nu, \tag{A1}$$

where $u^\mu$ is the four-velocity of the gas, $\rho$ is its rest mass density, $u$ is its internal energy density, and $b^\mu$ is a four-vector that describes the magnetic field in the fluid frame (Anile 1989; Komissarov 1999) with $b^\mu u_\mu = 0$ and $b^\mu b_\mu = b^2$. The gas pressure is $p_g = (\gamma_{\rm ad} - 1) u$, where $\gamma_{\rm ad} = 5/3$ is the adiabatic index. The temperature $T$ is defined in relativistic units such that $p_g = \rho T$, i.e., $T = (\gamma_{\rm ad} - 1)u/\rho$. Given the initial conditions of a system, the GRMHD code evolves the equation of mass conservation, $(\rho u^\mu)_{;\mu} = 0$, and the equations of energy-momentum conservation, $T^\mu_{\nu;\mu} = 0$, along with the ideal MHD induction equation. The output data describe the time evolution of the primitive quantities, $\rho$, $u$, $u^\mu$, and $b^\mu$, on each grid cell. At a given instant of time and at a given location in the grid, the quantity $\rho u^r \sqrt{-g}\, d\theta d\varphi$ measures the rate at which rest mass flows out radially within a solid angle of $d\theta d\varphi$, where $g = \det(g_{\mu\nu})$. The instantaneous net mass accretion rate through a sphere of radius $r$ is then

$$\dot{M}(r) \equiv - \iint \rho u^r \sqrt{-g}\, d\theta\, d\varphi, \quad (A2)$$

where the negative sign is because we associate accretion with inflow, which corresponds to a negative value of $u^r$. Because the accreting gas has large fluctuations in MHD (see the top panel in Figure 2), we are rarely interested in the instantaneous $\dot{M}$. Rather, we average over an extended period of time (e.g., the latter half of the entire simulation), and consider the time-averaged mass accretion rate, which we write as $\overline{\dot{M}}(r)$ with a bar. Furthermore, when we quote a unique accretion rate, we use the value at $r = 10\, r_g$, which we write as $\overline{\dot{M}}_{10}$. Sometimes it is useful to distinguish between mass that is flowing in at a given radius versus that flowing out (e.g., Figure 1). In this case, we separately define

$$\dot{M}_{\rm in}(r) \equiv \overline{- \iint \rho\, {\rm Min}(u^r, 0) \sqrt{-g}\, d\theta\, d\varphi},$$
$$\dot{M}_{\rm out}(r) \equiv \overline{- \iint \rho\, {\rm Max}(u^r, 0) \sqrt{-g}\, d\theta\, d\varphi}. \quad (A3)$$

Per this definition, $\dot{M}_{\rm in} \geqslant 0$ (inflow) and $\dot{M}_{\rm out} \leqslant 0$ (outflow). Following the convention used in this paper, $\dot{M}_{\rm in}$ and $\dot{M}_{\rm out}$ are shown in Figure 1 as solid lines (positive values) and dotted lines (negative values), respectively.

For the analyses in the paper, we define the time-averaged and $\rho$-weighted shell average of the quantity $X$ as

$$\langle X \rangle = \overline{\left( \frac{\iint X \rho \sqrt{-g}\, d\theta\, d\varphi}{\iint \rho \sqrt{-g}\, d\theta\, d\varphi} \right)}. \quad (A4)$$

The one exception is $\langle \rho \rangle$, which we compute as a straight shell average of the density without further $\rho$ weighting. In the shell averages, we leave out one closest cell from each of the two poles because these are affected by the polar boundary conditions.

For computing the flow of energy, we note that $T^r_t \sqrt{-g}\, d\theta d\varphi$ measures the rate at which energy *flows in* radially within a solid angle of $d\theta d\varphi$. Thus, the instantaneous energy inflow rate through a sphere at radius $r$ is given by

$$\dot{E}(r) \equiv \iint T^r_t \sqrt{-g}\, d\theta\, d\varphi. \quad (A5)$$

When energy flows toward the BH, $\dot{E}$ is positive. For studying feedback we are interested in the flow of energy away from the BH. We are also not interested in the flow of rest mass energy, which has no effect on the thermodynamics. We thus define the time-averaged net rate of outflow of *useful* energy as

$$\dot{E}_{\rm net}(r) \equiv \overline{(\dot{M} - \dot{E})} = - \iint \overline{(T^r_t + \rho u^r)} \sqrt{-g}\, d\theta\, d\varphi, \quad (A6)$$

and the corresponding efficiency $\eta^{\rm tot}$ as

$$\eta^{\rm tot}(r) = \dot{E}_{\rm net}(r)/\overline{\dot{M}}_{10} = \overline{(\dot{M} - \dot{E})}/\overline{\dot{M}}_{10}. \quad (A7)$$

To study the contribution of the fluid alone (ignoring the magnetic field), the relevant terms in the stress-energy tensor are

$$(T^{\rm fl})^r_t \equiv (\rho + \gamma_{\rm ad} u) u^r u_t. \quad (A8)$$

Then the time-averaged net fluid energy outflow rate is

$$\dot{E}^{\rm fl}_{\rm net}(r) \equiv - \iint \overline{(T^{\rm fl} + \rho u^r)^r_t} \sqrt{-g}\, d\theta\, d\varphi$$
$$= \iint \overline{Be\, \rho u^r} \sqrt{-g}\, d\theta\, d\varphi, \quad (A9)$$

where the relativistic Bernoulli parameter is (e.g., Penna et al. 2013, but not including the magnetic field)

$$Be = -(1 + \gamma_{\rm ad} u/\rho) u_t - 1. \quad (A10)$$

Following Ressler et al. (2021), we further subdivide the fluid energy outflow rate into an advective part,

$$\dot{E}^{\rm adv}_{\rm net}(r) \equiv \iint \overline{Be}\, \overline{\rho u^r} \sqrt{-g}\, d\theta\, d\varphi, \quad (A11)$$

and a convective part,

$$\dot{E}^{\rm conv}_{\rm net}(r) \equiv \dot{E}^{\rm fl}_{\rm net}(r) - \dot{E}^{\rm adv}_{\rm net}(r). \quad (A12)$$

In the same manner as in Equation (A7), we define the corresponding efficiencies, $\eta^{\rm fl}(r)$, $\eta^{\rm adv}(r)$, and $\eta^{\rm conv}(r)$.

In addition to the time-averaged and shell-integrated radial profiles $\dot{E}(r)$ defined above, we also define local time- and azimuth-averaged energy fluxes as a function of $r$ and $\theta$:

$$F^{\rm fl}_E(r, \theta) \equiv (1/2\pi) \int \overline{Be\, \rho u^r} \sqrt{-g}\, d\varphi, \quad (A13)$$

$$F^{\rm adv}_E(r, \theta) \equiv (1/2\pi) \int \overline{Be}\, \overline{\rho u^r} \sqrt{-g}\, d\varphi, \quad (A14)$$

$$F^{\rm conv}_E(r, \theta) \equiv F^{\rm fl}_E(r, \theta) - F^{\rm adv}_E(r, \theta). \quad (A15)$$

## Appendix B
## Numerical Setup

### B.1. HD Simulation Setup

In the hydrodynamic simulation described in Section 3, we use a domain with seven zones ($i = 0$–6) covering the range of $[1, 1.7 \times 10^7]\, r_g$. Zone 6 is initialized with a static fluid of constant density and temperature, and no perturbations, so as to recover the analytic Bondi solution without rotation. We use spherical Kerr–Schild coordinates with coordinate four-vectors $(x^t, x^r, x^\theta, x^\varphi)$, where the radial coordinate is $x^r \equiv \log r$, sometimes referred to as "eKS" coordinates. Rather than allow a vacuum to arise in the inner regions before the initial material from zone 6 accretes, new areas are filled with material before they are first simulated, with the same density and temperature as the innermost nonzero density zone present when they are initialized. Initialization in this way prevents disruption of the simulation, but bears no resemblance to the Bondi solution, validating that the scheme can converge to the correct result regardless of its initialization. After the first pass, no further initialization is needed. During the rest of the simulation, each active annulus receives half its initial data from the previous active annulus and half from the last output of the current active annulus. In the process, the inner and outer radial boundaries of the active annulus end up with new values of $\rho$, $u$, and $u^\mu$, which are treated as fixed boundary conditions while the annulus is evolved. Once the system has reached its final steady state, conservation of mass and energy between annuli are satisfied in a time-averaged sense.

**Table 1**
**Simulation Setup for Different Runs**

| Label | Resolution | Coordinates | Initial $\beta$ | Initial $\rho$ ($r < R_B$) |
|---|---|---|---|---|
| $128^3$, FMKS (fiducial) | $128^3$ | FMKS ($\chi_t = 0.8$, $\alpha = 16$) | $\sim 1$ | $r^{-1}$ |
| (i) $48^3$ | $48^3$ | MKS ($h = 0.3$) | $\sim 1$ | $r^{-1}$ |
| (ii) $64^3$ | $64^3$ | MKS ($h = 0.3$) | $\sim 1$ | $r^{-1}$ |
| (iii) $96^3$, FMKS | $96^3$ | FMKS ($\chi_t = 0.8$, $\alpha = 14$) | $\sim 1$ | $r^{-1}$ |
| (iv) $64^3$, $\beta 10$ | $64^3$ | MKS ($h = 0.3$) | $\sim 10$ | $r^{-1}$ |
| (v) $64^3$, const | $64^3$ | MKS ($h = 0.3$) | $\sim 1$ | Piecewise constant |
| (vi) $64^3$, Bondi | $64^3$ | MKS ($h = 0.3$) | $\sim 1$ | $r^{-3/2}$ |

### B.2. MHD Simulation Setup

The magnetized simulation in Section 4 is run over a larger domain with eight zones $[1, 1.3 \times 10^8] r_g$. Zone 7 is initialized with a similar density and temperature as those in the hydrodynamical model, but the two differ in several ways. The density is now initialized in the same way in all zones, to follow $\rho_{\text{init}}(r) = \rho_0 (r + R_B)/r$ such that it is $\propto r^{-1}$ interior to $R_B$ and constant outside of $R_B$. This is motivated by simulations (e.g., Ressler et al. 2020; Chatterjee & Narayan 2022; Guo et al. 2023) and observations (Chatterjee & Narayan 2022) that find a $\rho \propto r^{-1}$ relation in the presence of strong magnetic fields. However, the results are insensitive to the choice of $\rho_{\text{init}}$, as discussed in Appendix C. The temperature is initialized to the Bondi analytical solution, and the velocities are initialized to zero. The magnetic fields are initialized with a vector potential $A_\varphi(r, \theta) = b_z r (r + R_B) \sin^2 \theta / 2$ such that the plasma-$\beta$ parameter is approximately constant over all radii. The resulting magnetic field is vertical outside $R_B$, and becomes more radial inside $R_B$. The field is normalized with the parameter $b_z$ such that the initial $\beta_{\text{init}}(r) \sim 1$, motivated by observations of the Milky Way ISM (Ferrière 2020; Guerra et al. 2023). A per zone random perturbation of a maximum of 5% is applied to the internal energy when initializing fluid at all radii, to help break axisymmetry. Since the magnetic field picks out a rotation axis for the resulting flow parallel to the coordinate axis, we choose a coordinate system that compresses the $\theta$ coordinate away from coordinate axes and toward the midplane. We adopt a simplified version of *funky modified Kerr–Schild* (FMKS) coordinates (see, e.g., $\theta_j$ in the Appendix in Wong et al. 2021), eliminating the cylindrification term to obtain

$$\theta = Ny\left(1 + \frac{(y/\chi_t)^\alpha}{\alpha + 1}\right) + \frac{\pi}{2}, \quad (B1)$$

where $N \equiv \pi/2 (1 + \chi_t^{-\alpha}/(1 + \alpha))^{-1}$ is a normalization factor, $y = 2x^\theta - 1$, $\chi_t = 0.8$, $\alpha = 16$, and $x^\theta \in [0, 1]$. The simplified system retains the primary benefit of FMKS coordinates: zones near the coordinate pole are widened, relaxing the Courant condition and allowing a greater simulation time step, by a factor of 2 or more, dramatically reducing simulation cost. As we are interested primarily in steady-state behavior rather than resolving particular structures, e.g., a stable jet, the cylindrification is unnecessary for our case.

In addition to changing the coordinates, we lower the Courant factor for safety. We also introduce dynamical floors: in order to maintain $\rho > 10^{-6} r^{-3/2}$, internal energy $u > 10^{-8} r^{-5/2}$, temperature $u/\rho < 100$, and magnetization $b^2/\rho < 100$, we introduce material and energy in the coordinate frame. We also reduce the fluid velocity when the Lorentz factor measured by the Eulerian observer is larger than $\gamma_{\text{max}} = 10$. While running each active annulus, we keep $\rho$, $u$, and $u^\mu$ fixed at the inner and outer boundaries, as in the hydro case. In addition, we also keep $b^\mu$ fixed and make sure that we preserve $\nabla \cdot B = 0$.

We have validated the *multi-zone* method described in Section 2 in the presence of strong magnetic fields, by comparing a smaller four-zone simulation with $R_B \approx 460$ against a single-zone simulation of the same problem. We obtained substantially similar averaged radial profiles. These comparisons, along with detailed descriptions of the method, boundary conditions, etc., will be presented in a forthcoming longer paper.

### Appendix C
### Study of Simulation Resolutions and Initial Conditions

We have presented our fiducial MHD simulation results in Section 4 for a resolution of $128^3$ per annulus with the FMKS coordinate system and initialized with $\rho_{\text{init}} \propto r^{-1}$ at $r < R_B$. Here we show that the results remain the same even when we choose different resolutions, coordinate systems, or initial conditions. First, we introduce a new modified Kerr–Schild (MKS) coordinate system (see, e.g., $\theta_g$ in the Appendix of Wong et al. 2021) where the resolution is more focused in the midplane but the zones near the pole are not widened:

$$\theta = \pi x^\theta + \frac{1}{2}(1 - h)\sin(2\pi x^\theta). \quad (C1)$$

Here we use $h = 0.3$. In the case of $h = 1$, as it is equivalent to the uniform $\theta$ grid of the eKS coordinates in Appendix B.1.

Six extra simulations were run with parameters outlined in Table 1. Using MKS coordinates and other setup details that are the same as those in our fiducial run, we ran simulations at (i) a $48^3$ resolution per annulus and (ii) a $64^3$ resolution per annulus. Run (iii) with a $96^3$ resolution per annulus used FMKS but with a different $\alpha = 14$, which is less extreme than $\alpha = 16$ in the fiducial run. Run (iv) used a weaker initial magnetic field with $\beta \sim 10$ and used MKS coordinates with a $64^3$ resolution per annulus. Runs (v) and (vi) were initialized with a different initial density profile from the other runs and used MKS with a $64^3$ resolution. Run (v) was initialized with a piecewise constant density profile as in the HD run. Run (vi) was initialized with the analytical HD Bondi solution where the density scales as $\rho_{\rm init} \propto r^{-3/2}$ for $r < R_{\rm B}$.

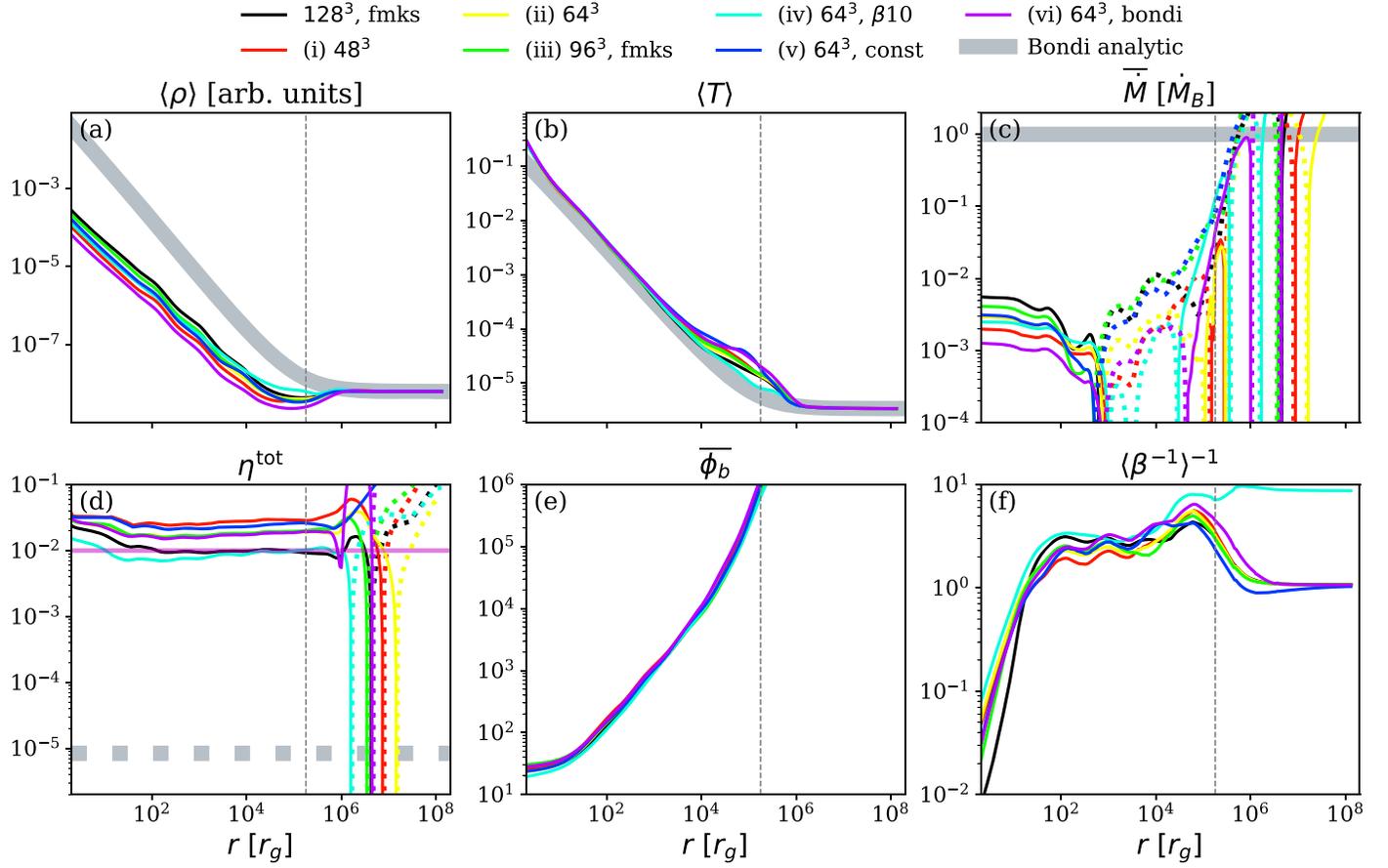

**Figure 4.** Comparison between simulations with varying resolutions, coordinate systems, and initial magnetic strength, described in Table 1. Each profile is averaged between $t_{\rm run}(R_{\rm B}) = 4.5\, t_{\rm B} - 9\, t_{\rm B}$ measured at the Bondi radius (gray vertical dashed line). Positive (negative) values are shown in solid (dotted) lines. Overall, we find good agreement between these simulations.

Radial profiles of all the quantities of interest for the six extra simulations are compared with the fiducial run in Figure 4. Overall, the profiles are very similar between the various runs implying that our conclusions from the fiducial run hold true regardless of the choice of coordinate system, resolution, initial magnetic field strength, or initial density profile. The accretion rate at $10\, r_g$ of all five runs lies in the range of $\overline{\dot{M}} \approx (1-5) \times 10^{-3}\, \dot{M}_{\rm B}$, which is consistent with observations of M87* (Russell et al. 2015; Event Horizon Telescope Collaboration et al. 2021) and Sgr A* (Wang et al. 2013). The outflow efficiency is around $\eta^{\rm tot} \approx 1\%\text{--}3\%$, where the highest value is from the lowest-resolution ($48^3$) run (i). Even though initialized with a weaker magnetic field, run (iv) still shows magnetic field saturation at $\beta \sim 3$ at radii $r = 10^2\text{--}10^4$. Near the horizon, run (iv) accreted the least amount of magnetic field as can be inferred from its larger $\beta$ and smaller $\phi_b$. However, the difference is not large. The $\beta$-parameter decreases more steeply toward the BH for run (iii) and the fiducial run, both of which use FMKS coordinates. We suspect this is because the FMKS coordinates are characterized by larger cells near the poles where $\beta^{-1}$ tends to be larger, and this might have affected the shell average. Runs (v) and (vi), despite their very different initial density profiles, converge to a similar final density profile with $\rho \propto r^{-1}$. This demonstrates that our results are not dependent on the initial conditions.

**ORCID iDs**

Hyerin Cho (조혜린) ⓘ https://orcid.org/0000-0002-2858-9481
Ben S. Prather ⓘ https://orcid.org/0000-0002-0393-7734
Ramesh Narayan ⓘ https://orcid.org/0000-0002-1919-2730
Priyamvada Natarajan ⓘ https://orcid.org/0000-0002-5554-8896
Kung-Yi Su ⓘ https://orcid.org/0000-0003-1598-0083
Angelo Ricarte ⓘ https://orcid.org/0000-0001-5287-0452
Koushik Chatterjee ⓘ https://orcid.org/0000-0002-2825-3590

**References**

Anglés-Alcázar, D., Quataert, E., Hopkins, P. F., et al. 2021, ApJ, 917, 53
Anile, A. M. 1989, Relativistic fluids and magneto-fluids: With applications in astrophysics and plasma physics (Cambridge: Cambridge Univ. Press)
Bisnovatyi-Kogan, G. S., & Ruzmaikin, A. A. 1974, Ap&SS, 28, 45
Bondi, H. 1952, MNRAS, 112, 195
Booth, C. M., & Schaye, J. 2009, MNRAS, 398, 53
Chatterjee, K., Chael, A., Tiede, P., et al. 2023, Galax, 11, 38
Chatterjee, K., & Narayan, R. 2022, ApJ, 941, 30
Davé, R., Anglés-Alcázar, D., Narayanan, D., et al. 2019, MNRAS, 486, 2827
Event Horizon Telescope Collaboration, Akiyama, K., Algaba, J. C., et al. 2021, ApJL, 910, L13
Ferrarese, L., & Merritt, D. 2000, ApJL, 539, L9
Ferrière, K. 2020, PPCF, 62, 014014
Fiacconi, D., Sijacki, D., & Pringle, J. E. 2018, MNRAS, 477, 3807
Gammie, C. F., McKinney, J. C., & Tóth, G. 2003, ApJ, 589, 444
Gebhardt, K., Bender, R., Bower, G., et al. 2000, ApJL, 539, L13
Grete, P., Dolence, J. C., Miller, J. M., et al. 2023, Int. J. High Perform. Comput. Appl., 37, 465

Guerra, J. A., Lopez-Rodriguez, E., Chuss, D. T., Butterfield, N. O., & Schmelz, J. T. 2023, AJ, 166, 37
Guo, M., Stone, J. M., Kim, C.-G., & Quataert, E. 2023, ApJ, 946, 26
Hopkins, P. F., Grudic, M. Y., Su, K.-Y., et al. 2023, arXiv:2309.13115
Hopkins, P. F., & Quataert, E. 2010, MNRAS, 407, 1529
Igumenshchev, I. V., & Narayan, R. 2002, ApJ, 566, 137
Igumenshchev, I. V., Narayan, R., & Abramowicz, M. A. 2003, ApJ, 592, 1042
Kaaz, N., Murguia-Berthier, A., Chatterjee, K., Liska, M. T. P., & Tchekhovskoy, A. 2023, ApJ, 950, 31
Komissarov, S. S. 1999, MNRAS, 303, 343
Kormendy, J., & Ho, L. C. 2013, ARA&A, 51, 511
Lalakos, A., Gottlieb, O., Kaaz, N., et al. 2022, ApJL, 936, L5
Li, Y., & Bryan, G. L. 2014, ApJ, 789, 153
Magorrian, J., Tremaine, S., Richstone, D., et al. 1998, AJ, 115, 2285
Meszaros, P. 1975, A&A, 44, 59
Michel, F. C. 1972, Ap&SS, 15, 153
Narayan, R., Chael, A., Chatterjee, K., Ricarte, A., & Curd, B. 2022, MNRAS, 511, 3795
Narayan, R., Igumenshchev, I. V., & Abramowicz, M. A. 2000, ApJ, 539, 798
Narayan, R., Igumenshchev, I. V., & Abramowicz, M. A. 2003, PASJ, 55, L69
Ni, Y., Di Matteo, T., Bird, S., et al. 2022, MNRAS, 513, 670
Pen, U.-L., Matzner, C. D., & Wong, S. 2003, ApJL, 596, L207
Penna, R. F., Kulkarni, A., & Narayan, R. 2013, A&A, 559, A116
Porth, O., Chatterjee, K., Narayan, R., et al. 2019, ApJS, 243, 26
Prather, B. S., Wong, G. N., Dhruv, V., et al. 2021, JOSS, 6, 3336
Quataert, E., & Gruzinov, A. 2000, ApJ, 539, 809
Ressler, S. M., Quataert, E., White, C. J., & Blaes, O. 2021, MNRAS, 504, 6076
Ressler, S. M., White, C. J., Quataert, E., & Stone, J. M. 2020, ApJL, 896, L6
Ricarte, A., Tremmel, M., Natarajan, P., & Quinn, T. 2019, MNRAS, 489, 802
Ripperda, B., Liska, M., Chatterjee, K., et al. 2022, ApJL, 924, L32
Rosas-Guevara, Y., Bower, R. G., Schaye, J., et al. 2016, MNRAS, 462, 190
Russell, H. R., Fabian, A. C., McNamara, B. R., & Broderick, A. E. 2015, MNRAS, 451, 588
Shapiro, S. L., & Teukolsky, S. A. 1983, Black Holes, White Dwarfs, and Neutron Stars : the Physics of Compact Objects (New York: Wiley)
Shvartsman, V. F. 1971, SvA, 15, 377
Sijacki, D., Vogelsberger, M., Genel, S., et al. 2015, MNRAS, 452, 575
Talbot, R. Y., Bourne, M. A., & Sijacki, D. 2021, MNRAS, 504, 3619
Tchekhovskoy, A., McKinney, J. C., & Narayan, R. 2012, JPhCS, 372, 012040
Tchekhovskoy, A., Narayan, R., & McKinney, J. C. 2011, MNRAS, 418, L79
Trott, C. R., Lebrun-Grandi, D., Arndt, D., et al. 2022, IEEE Transactions on Parallel & Distributed Systems, 33, 805
Wang, Q. D., Nowak, M. A., Markoff, S. B., et al. 2013, Sci, 341, 981
Weinberger, R., Springel, V., Pakmor, R., et al. 2018, MNRAS, 479, 4056
Weinberger, R., Su, K.-Y., Ehlert, K., et al. 2023, MNRAS, 523, 1104
Wellons, S., Faucher-Giguère, C.-A., Hopkins, P. F., et al. 2023, MNRAS, 520, 5394
Wong, G. N., Du, Y., Prather, B. S., & Gammie, C. F. 2021, ApJ, 914, 55
Yuan, F., Gan, Z., Narayan, R., et al. 2015, ApJ, 804, 101